\documentstyle[11pt,epsfig]{article}

\def\roughly#1{\mathrel{\raise.3ex\hbox{$#1$\kern-.75em%
\lower1ex\hbox{$\sim$}}}}
\def\fm{{\rm fm}}
\def\lsim{\roughly<}

\def\be{\begin{eqnarray}}
\def\ee{\end{eqnarray}}
\def\Tr{{\rm Tr}\;}

\def\ben{\begin{enumerate}}
\def\een{\end{enumerate}}
\def\beitem{\begin{itemize}}
\def\eitem{\end{itemize}}
\def\fm{{\rm fm}}

\newcommand{\beq}{\begin{eqnarray}}
\newcommand{\eeq}{\end{eqnarray}}

\def\bi{\begin{itemize}}
\def\ei{\end{itemize}}

\def\etal{{\it et al}}
\def\del{\partial}
\def\L{{\cal L}}

\long\def\beginomit#1\endomit{}
\def\np{{\it Nucl. Phys.}}
\def\prl{{\it Phys. Rev. Lett.}}

\def\pl{{\it Phys. Lett.}}
\def\L{{\cal L}}

\def\AJ{{\it Astrophys. J.}}
\def\AA{{\it Astron. \& Astrophys.}}
\def\PR{{\it Phys. Repts.}}
\def\chpt{$\chi$PT}
\def\be{\begin{eqnarray}}
\def\ee{\end{eqnarray}}
\def\Tr{{\rm Tr}\;}

\def\fm{{\rm fm}}
\def\etal{{\it et al.}}

\def\del{\partial}
\def\L{{\cal L}}
\def\M{{\cal M}}
\def\MeV{{\mbox MeV}}
\def\M{{\cal M}}

\begin{document}

%-------------------------------------------------------------------------%
%                     Title Page                                          %
%-------------------------------------------------------------------------%

\begin{titlepage}\begin{center}

\hfill{T94/xxx, SNUTP-94-121}

%\hfill{hep-ph/9411364}

\hfill{November 1994}

\vskip 0.4in
{\large\bf KAON CONDENSATION IN DENSE STELLAR MATTER\footnote{Based on
talk given by MR at {\it International Symposium on ``Strangeness and
Quark Matter"}, September 1--5, 1994, Krete, Greece and at {\it
YITP Workshop ``From Hadronic Matter to Quark Matter: Evolving View of
Hadronic Matter"}, October 30--November 1, 1994, Yukawa Institute, Kyoto
University, Kyoto, Japan.}}
\vskip 0.6in
{\large  Chang-Hwan Lee$^a$ and Mannque Rho$^b$}\\
\vskip 0.1in
{\large a) \it Department of Physics and Center for Theoretical Physics} \\
{\large \it Seoul National University, Seoul 151-742, Korea}\\
{\large b) \it Service de Physique Th\'{e}orique, CEA  Saclay}\\
{\large\it 91191 Gif-sur-Yvette Cedex, France}\\
\vskip .6in
%\centerline{November 1994}
\vskip .6in

{\bf ABSTRACT}\\ \vskip 0.1in
\begin{quotation}

\noindent This talk is based on work done in collaboration with
G.E. Brown and D.-P. Min on kaon condensation in dense baryonic medium treated
in chiral perturbation theory using heavy-baryon formalism. It contains,
in addition to what was recently published, some new results based on
the analysis on kaonic atoms by Friedman, Gal
and Batty and a discussion on a renormalization-group analysis
to meson condensation made together with H.K. Lee and Sin. Negatively charged
kaons are predicted to condense at the
critical density $2\lsim \rho/\rho_0\lsim 4$, in the range to allow all the
intriguing new phenomena predicted by Brown and Bethe to take place
in compact star matter.
\end{quotation}
\end{center}\end{titlepage}
\section{Motivation}

Recent work by Bethe and Brown\cite{BB} on the maximum mass of stable
compact stars --
called ``neutron stars" in the past but more appropriately
``nuclear (or nucleon) stars" -- suggest that the nuclear equation of state
(EOS) in the interior of compact stars must be considerably softened
at densities a few times the nuclear matter density $\rho_0$ by one or
several hadronic phase transitions. It is now fairly clear that
neither pion condensation nor quark matter will figure at a density low enough
to be relevant to the star matter although the issue is not yet
completely settled.
As Bethe and Brown suggest, kaon condensation could however take place
at a density 3--4 times the normal matter density and hence play an
important role in explaining the remarkably
narrow range of compact star masses observed in nature\cite{mass}.

The aim of this talk is to describe an honest calculation that predicts
the critical density for kaon condensation. The strategy is to take up
what Kaplan and Nelson\cite{KN} started, namely chiral perturbation
theory (\chpt). \ Kaplan and Nelson predicted in tree order of \chpt \ that
kaons condense in neutron matter at $\rho\lsim 3\rho_0$. Our calculation
goes to next-to-next-to-leading (NNL) order. It turns out that
the calculation confirms the Kaplan-Nelson prediction although in the
process new and interesting physical elements are uncovered. Our result is that
for reasonable ranges of parameters involved, the critical density
comes out to be
\be
2\lsim \rho/\rho_0 \lsim 4.
\ee
This is the range of densities relevant to the Bethe-Brown scenario for the
formation of light-mass black holes for stars that exceed the
critical mass of $M=1.5 M_\odot$.

\section{Where Do the Kaons Come From?}
There are two situations where the production of kaons brings out
interesting physics. One is their properties in relativistic heavy-ion
collisions that involve temperature. Here kaon condensation is
not directly relevant but the {\it mechanism} that triggers kaon condensation
in the relevant situation has intriguing consequences on the properties
of kaons observed in heavy-ion experiments. This is discussed in a recent
review\cite{newbr94} and will not be discussed here. What we are interested
in is what kaons do in cold dense matter appropriate to compact objects
that result from the collapse of massive stars.

In stellar collapse, as matter density $\rho$ increases, the electron
chemical potential $\mu_e$ (determined by the chemical potentials of neutrons
and protons in the system together with charge neutrality)
increases, reaching several hundreds of MeV.
If the electron chemical potential reaches the ``effective mass" of
a meson $\Phi$, $m_\Phi$, then the electron can ``decay" into a $\Phi$
as\cite{BKR}
\be
e^-\rightarrow \Phi^- +\nu_e.\label{edecay}
\ee
In nature, the only low-mass bosons are the pseudo-Goldstone bosons
$\Phi^-=\pi^-, K^-$. While lowest in mass, the pions do not seem to play
an important role, so the next possible boson is the kaon with
its mass $\sim 500$ MeV in free space. The electron chemical potential
cannot reach this high, so on-shell kaons cannot be produced by this process.
However as will be described below, the kaon in medium can undergo a mass
shift due to density-dependent renormalization. As the $\mu_e$ increases
and the effective kaon mass
$M_K^\star$ decreases as $\rho$ increases, the process (\ref{edecay})
can occur at some density $\rho_c$. Kaons so produced will bose-condense
at that density $\rho_c$. Whether or not this will occur then depends on
whether or not $M_K^\star$ will decrease enough in density so that
it meets $\mu_e$. Such a condensation will be of physical interest if
the critical density is low enough and the energy gain is high enough.
This is the possibility we shall address below.

\section{Baryon Chiral Perturbation Theory}
\subsection{\it Effective Field Theory for Nuclear Matter}
The process we are interested in requires a field theory that can
describe simultaneously normal nuclear matter and phase transitions therefrom.
The most relevant ingredient of QCD that is needed here is
spontaneously broken chiral symmetry. We are specifically interested
in chiral $SU(3)\times SU(3)$ symmetry since strangeness is involved.
In order to address the problem, we need to start from a realistic
effective chiral Lagrangian, obtain a nuclear matter of the right
properties from it and then determine whether strangeness condensation
occurs.

Unfortunately we do not yet know how to describe nuclear matter starting from
a chiral Lagrangian. There are various suggestions and one promising one is
that nuclear matter arises as a solitonic matter from a chiral effective
action, a sort of chiral liquid\cite{lynn} resembling Landau Fermi liquid.
The hope is that the resulting effective action would look like
Walecka's mean-field model. There is as yet no convincing derivation along
this line. In the work reported here, we will have to assume that we have
a nuclear matter  that comes out of an effective chiral action. Given
such a ground state containing no strange degrees of freedom, we would
like to study fluctuations along the strangeness direction and
determine if instability along that direction develops
signaling a phase transition.
We are therefore assuming that we can get the properties of normal nuclear
matter from phenomenology, that is, that nuclear matter is a
Fermi-liquid fixed point\cite{shankar,polchinski}. In principle,
a precise knowledge
of this ground state from a chiral effective Lagrangian at a nonperturbative
QCD level would
allow us to determine the coefficients that appear in the effective Lagrangian
with which to describe fluctuations
around the soliton background -- i.e.,
the Fermi liquid --and with which we could then compute all nuclear response
functions. At present such a derivation does not exist. In a recent paper
by Brown and one of the authors (BR91)\cite{brscaling},
it is assumed that in medium at a matter density $\rho\sim \rho_0$,
the {\it nuclear} effective field theory can be written in terms of
the medium-dependent coupling constants $g^\star$ and masses
of hadrons $m^\star$ while preserving the free-space structure of
a sigma model. This leads to the so-called Brown-Rho scaling. In
BR91\cite{brscaling}, the nonlinear sigma model implemented with trace anomaly
of QCD is used to arrive at the scaling law. The precise way that this
scaling makes sense is elaborated by Adami and Brown\cite{adamibrown} and
in the review (BR94)\cite{newbr94}. There have been numerous papers written
with some of the essential points of this scaling misinterpreted.

Given such an effective field theory, we can make a general argument on the
stability in various flavor directions of nuclear matter at high density.
This can be done along the line of arguments developed for condensed matter
physics by
Shankar\cite{shankar} and Polchinski\cite{polchinski} using renormalization
group flow. We sketch the essential argument following Lee, Rho and
Sin\cite{LRS}.

What we are interested in is whether the system in question develops
instability along the direction of strangeness and if so, by which
physical mechanism. This analysis will not give us the critical density.
The critical density will be calculated by using chiral perturbation
theory. For this purpose we will focus on the kaon frequency near
the electron chemical potential. By Baym's theorem\cite{baym},
one can identify
the kaon chemical potential associated with charge conservation,
$\mu_K$, with the electron chemical potential, $\mu_e$, which we shall simply
write $\mu$ in what follows. This means that we will be looking
at the vicinity of $\omega\sim \mu$ in the kaon dispersion formula.
We shall assume that
\be
|\omega-\mu| << \mu.
\ee
As mentioned, we assume
that nucleons in nuclear matter are in Fermi-liquid state with the Fermi
energy $\mu_F$ and the Fermi momentum $k_F$.  Define $\psi$ as the nucleon
field fluctuating around the Fermi surface
such that the momentum integral has a cut-off $\Lambda_N$,
\be
k_F -\Lambda_N < |\vec{k}| < k_F + \Lambda_N.
\ee
Kaons can interact with the nucleons through three-point functions of
the $KNN$ type (Yukawa interaction) and through four-point interactions
of the $KKNN$ type. We shall consider s-wave kaon-nucleon interactions,
for which the Yukawa interaction can be ignored. A generic action involving
the nucleon field $\psi$ and the kaon field
$\Phi$ can then be written, schematically, as
\be
S &=&  \int d\omega  d^3q \Phi^{*}(\omega, \vec{q})\left(\omega -
q^2/2\mu_K\right)\Phi(\omega, \vec{q})-\int d\omega  d^3q\, \tilde{M}_K
\Phi^*\Phi \nonumber\\
&& + \int (d\omega d^3q)^2 (d\epsilon d^3k)^2 h \Phi^*\Phi \psi^{\dagger}
 \psi \delta^4(\omega,\epsilon, \vec{q}, \vec{k})\nonumber\\
&& + \int d\epsilon d^3k \psi^{\dagger}\left(\epsilon -\epsilon(k))\right)\psi
    +g \int (d\epsilon d^3k)^4 \psi^{\dagger}\psi^{\dagger}\psi\psi
    \delta^4(\epsilon, \vec{k})
\label{toy1}
\ee
where $\tilde{M}=(M_K^2-\mu^2)/2\mu$ and $h$ and $g$ are constants. The
four-Fermi interaction with the coefficient $g$ stands for Fermi-liquid
interactions in nuclear matter. (In nuclear matter, one can have four such
terms because of the nucleon spin and isospin degrees of freedom. We need
not specify
them for our purpose.) This is a toy action but it is generic in that the
results of \chpt \ we will obtain below can be put into this form.

The renormalization group flow of this action can be analyzed in the
following way. Since we are assuming that nuclear matter is a Fermi-liquid
fixed point, fluctuations in the non-strange direction in the nucleon sector
are stable: The four-Fermi interaction $g$ is irrelevant or at best marginal.
Fluctuations in the strange direction involve the kaon field $\Phi$.
Suppose we have integrated out all the high-frequency modes above the
cut-off $\Lambda$ measured with respect to $\mu$. We are interested in the
stability of the system
under the renormalization group transformation $\Lambda\rightarrow s\Lambda$
($s<1$) as $s\rightarrow 0$. A scaling analysis shows that the
interaction term $h$ is
irrelevant while the ``mass term" $\tilde{M}$ is relevant. The renormalization
group-flow of the ``mass term" and the interaction term $h$ can be
readily written down and solved\cite{LRS} (with $t=-\ln\ s$),
\be
\tilde{M}(t)=(\tilde{M}_0-{Dh_0\over 1+a})e^{t} + {Dh_0\over 1+a}e^{-at}
\label{sol}
\ee
with
\be
h(t)=h_0 e^{-at},\ \ \ h_0\geq 0
\ee
where $D=\frac{3(1+\alpha^2)\alpha}{2\mu}\rho_N >0$, $\alpha= \Lambda/k_F >0$
and $a=1/2$. We see from Eq.(\ref{sol}) that as $s\rightarrow 0$ for which
$h\rightarrow 0$,  $\tilde{M}$ changes sign for some
$(\tilde{M}_0, h_0\geq 0)$. Thus although irrelevant, an attractive interaction
$h_0$ determines the direction of the mass flow whereas it is the ``mass term"
that drives the system to instability.

\subsection{\it Chiral Counting}
Armed with the general information on the instability in the strangeness
direction, we now calculate the critical density in \chpt.\ Remember that
we are to look at the instability in the kaon direction, so it suffices
for us to look at the fluctuations around the Fermi-liquid state.
For this we need an effective chiral Lagrangian involving baryons as well
as Goldstone bosons. When baryons are present, \chpt \ is not as firmly
formulated as when they are absent\cite{leutwyler}. The reason is that
the baryon mass $m_B$ is $\sim \Lambda_\chi\sim 1$ GeV, the chiral
symmetry breaking scale. It is more expedient, therefore, to redefine
the baryon field so as to remove the mass from the baryon propagator
\be
B_v=e^{im_B\gamma\cdot v\; v\cdot x} P_+ B
\ee
where $P_+=(1+\gamma\cdot v)/2$ and write the baryon four-momentum
\be
p_\mu=m_B v_\mu +k_\mu
\ee
where $k_\mu$ is the small residual momentum indicating the baryon
being slightly off-shell.
When acted on by a derivative, the baryon field $B_v$ yields
a term of $O(k)$. Chiral perturbation theory in terms of $B_v$ and
Goldstone bosons $(\pi\cdot\lambda/2)$ is known as
``heavy-baryon (HB) \chpt"\cite{HFF}. HB\chpt \ consists of making
chiral expansion in derivatives on Goldstone boson fields,
$\del_M/\Lambda_\chi$,
and on baryon fields, $\del_B/m_B$,  and in the quark mass matrix,
$\kappa {{\cal M}}/\Lambda_\chi^2$. In the meson sector, this is just what
Gasser and Leutwyler did for $\pi\pi$ scattering. In the baryon sector,
consistency with this expansion requires that the chiral counting be made
with $B^\dagger (\cdots)B$, not with $\bar{B} (\cdots)B$. This means that
in medium, it is always the baryon density $\rho (r)$ that comes in and
{\it not} the scalar density $\rho_s (r)$. This point seems to be
misunderstood by some workers in the field, a cause for one of the
red herrings in the literature.

Following Weinberg\cite{weinberg}, we organize the chiral expansion
in power $Q^\nu$ where $Q$ is the characteristic energy/momentum scale
we are looking at ($Q<< \Lambda_\chi$) and
\be
\nu=4-N_n-2C+2L +\sum_i \Delta_i
\ee
with the sum over $i$ running over the vertices that appear in the graph
and
\be
\Delta_i=d_i +\frac 12 n_i -2.
\ee
Here $\nu$ gives the power of small momentum (or energy) for a process
involving $N_n$ nucleon lines, %$N_K$ kaon lines,
$L$ number of loops,
$d_i$ number of derivatives (or powers of meson mass) in the $i$th
vertex, $n_i$ number of nucleon lines entering into $i$th vertex and
$C$ is the number of separate connected pieces of the Feynman graph.
Chiral invariance requires that $\Delta_i\geq 0$, so that the leading
power is given by $L=0$, $\nu=4-N_N-2C$.

As an example, consider $KN$ scattering.
The leading term here is the tree graph with $\nu=1$ and with $N_n=C=1$.
The next order terms are $\nu=2$ tree graphs with $\Delta=1$ that involves
two derivatives or one factor of the mass matrix ${\cal M}$.
{}From $\nu=3$ on, we have loop graphs contributing
together with appropriate counter terms.

In considering kaon-nuclear interactions as in the case of kaon condensation,
we need to consider the case with $N_n\geq 2$ and $C\geq 2$. In dealing
with many-body system, one can simply fix $4-N_n$ and consider $C$ explicitly.
For instance if one has two nucleons (for reasons mentioned below, this
is sufficient, with multinucleon interactions being suppressed), then
we have $4-N_n=2$ but $C$ can be 2 or 1, the former describing a kaon
scattering on a single nucleon with a spectator nucleon propagating
without interactions and the latter a kaon scattering irreducibly on a
two-nucleon complex.  Thus intrinsic $n$-nucleon processes are suppressed
compared with $(n-1)$-nucleon processes by at least $O(Q^2)$.
This observation will be used later for arguing that four-Fermi interactions
are negligible in kaon condensation.  This is somewhat like the suppression
of three-body nuclear forces\cite{weinberg} and of three-body exchange
currents\cite{PMR} in chiral Lagrangians.

\subsection{\it Kaon-Nucleon Scattering}

Given a chiral Lagrangian, we need to first determine the parameters of the
Lagrangian from available phenomenology.
This is inevitable in effective field theories. We shall
first look at kaon-nucleon scattering at low energies. This was done
by Lee {\etal}\cite{LJMR,LBMR} which we summarize here. We shall compute
the scattering amplitude to one-loop order and this entails a Lagrangian
written to $O(Q^3)$ as one can see from the Weinberg counting rule.
Instead of writing it out in its full glory, we write it in a schematic
form as
\be
\L=\sum_i \L_i [B_v, U, \M]\label{lag}
\ee
where the subscript $i$ stands for $\nu$ relevant to the $KN$ channel.
Here $B_v$ stands for both octet and decuplet baryons and $U$ the Sugawara
form for octet Goldstone bosons.  For $KN$ scattering in free-space,
the Lagrangian is bilinear in the baryon field.
Details are given in Lee {\etal} \cite{LJMR,LBMR}.
Let us specify a few terms in (\ref{lag}) so as to streamline our discussion.
Focusing on s-wave scattering, $\L_1$ contains the leading order term that may
be described by the exchange of an $\omega$ between kaon and nucleon,
attractive for $K^-N$ and repulsive for $K^+ N$ and an isovector term
corresponding to the exchange of a $\rho$ meson. These terms are proportional
to the kaon frequency $\omega$. To next order,
$\L_2$ contains the ``$KN$ sigma term" proportional to $\Sigma_{KN}/f^2$
where $f$ is the pion decay constant and a term proportional to $\omega^2$
which may be saturated by decuplet intermediate states. The $\nu=3$ pieces are
counter terms that contain terms that remove divergences in the loop
calculations and finite terms that are to be determined from experiments.
The complete s-wave scattering amplitudes calculated
to the NNL order come out to be
\be
    a_0^{K^\pm p} &=& \frac{m_B}{4\pi f^2 (m_B+M_K)}
    \left[
    \mp M_K
    + (\bar d_s+\bar d_v) M_K^2 +\left\{ (L_s+L_v) \pm (\bar g_s +\bar g_v)
    \right\} M_K^3 \right]
  \nonumber\\
&&+\delta a_{\Lambda^\star}^{K^\pm p}\nonumber\\
  a_0^{K^\pm n} &=& \frac{m_B}{4\pi f^2 (m_B+M_K)}
    \left[ \mp \frac 12 M_K
    + (\bar d_s-\bar d_v) M_K^2 +\left\{ (L_s-L_v) \pm (\bar g_s -\bar g_v)
    \right\} M_K^3 \right]\nonumber\\
  \label{scattamp}\ee
  where $M_K$ is the kaon mass, $m_B$ the baryon (nucleon) mass,
$\bar d_s$ is the t-channel isoscalar contribution [${\cal O}(Q^2)$], and
$\bar d_v$ is the t-channel isovector one [${\cal O}(Q^2)$], both coming
from $\L_2$,  $L_s$($L_v$) is the finite crossing-even t-channel isoscalar
(isovector) finite one-loop contribution [$O(Q^3)$]
having the numerical values
  \be
    L_s M_K \approx -0.109 \; \fm, \ \ \ \
   L_v M_K    \approx +0.021 \; \fm
   \ee
and the quantity $\bar g_s (\bar g_v)$ is the crossing-odd t-channel isoscalar
(isovector) contribution [$O(Q^3)$] from one-loop plus counter terms in
$\L_3$. Finally $\delta a_{\Lambda^\star}^{K^\pm p}$ is the contribution from
the $\Lambda (1405)$ intermediate state Born diagram,
\be
   \delta a_{\Lambda^\star}^{K^\pm p} &=& - \frac{m_B}{4\pi f^2 (m_B+M_K)}
   \left[ \frac{ g_{\Lambda^\star}^2 M_K^2}{m_B \mp M_K -
m_{\Lambda^\star}}\right]
   \label{Lama}\ee
which is completely determined given experimental data on the coupling
  $ g_{\Lambda^\star}$ and the complex mass $m_{\Lambda^\star}$.

There are four unknowns $\bar{d}_{s,v}$, $\bar{g}_{s,v}$ in (\ref{scattamp})
which can be determined from four experimental (real part of)
scattering lengths
  \be
  a_0^{K^+p} = -0.31 \fm,& \;\;\;\;& a_0^{K^-p} = -0.67 +i 0.63 \fm
  \nonumber\\
  a_0^{K^+n} = -0.20 \fm,& \;\;\;\;& a_0^{K^-n} = +0.37 +i 0.57 \fm .
\label{expscatt}
  \ee
The results are
   \be
   \bar d_s \approx 0.201  \fm, \;\; && \bar d_v \approx 0.013 \fm,
   \nonumber\\
   \bar g_s M_K \approx 0.008 \fm, \;\; && \bar g_v M_K \approx 0.002 \fm.
   \label{num}\ee

So far, no prediction is made. However given the parameters so fixed,
one can then go ahead and calculate the s-wave amplitude that enters
in kaon condensation. This amounts to going off-shell in the $\omega$
variable, that is, in the kinematics where $\omega\neq M_K$. In doing this,
one encounters an ambiguity due to the $\omega$ dependence of the
coefficients $\bar{d}$ which consist of the ``$KN$ sigma term" and ``$\omega^2$
term" which get compounded on-shell into one term. In the calculation
reported in Lee {\etal}\cite{LJMR,LBMR}, we chose to fix the ``$\omega^2$
term" by resonance saturation and leave the ``sigma term" to be fixed
by the on-shell data. The predicted off-shell amplitudes\cite{LJMR}
agree reasonably with phenomenologically constructed off-shell amplitudes.
All the constants of the chiral Lagrangian bilinear in the baryon field
are thereby determined to $O(Q^3)$.

\subsection{\it Four-Fermi Interactions}

In medium, the chiral Lagrangian can have multi-Fermi interactions
as a result of ``mode elimination." Here we consider four-Fermi interactions,
ignoring higher-body interactions. We shall see that this is justified.

As stated above, we need to focus on four-Fermi interactions that involve
strangeness degrees of freedom. Nonstrange four-Fermi interactions are subsumed
in the Fermi-liquid structure of normal nuclear matter. For s-wave kaon-nuclear
interactions, we only have the $\Lambda (1405)$ to account for. There are
only two terms,
 \be
    {\cal L}_{4-fermion} &=&
   C_{\Lambda^\star}^S \bar{\Lambda}^\star_v \Lambda^\star_v
  \Tr \bar B_v B_v +  C_{\Lambda^\star}^T
  \bar{\Lambda}^\star_v \sigma^k \Lambda^\star_v
     \Tr \bar B_v \sigma^k B_v\label{fourfermi}
    \ee
where $C_{\Lambda^\star}^{S,T}$ are the
dimension $-2$ ($M^{-2}$) parameters to be fixed
empirically and $\sigma^k$ acts on baryon spinor. We shall now describe
how to fix these two parameters from kaonic atom data.

\begin{figure}
 \centerline{\epsfig{file=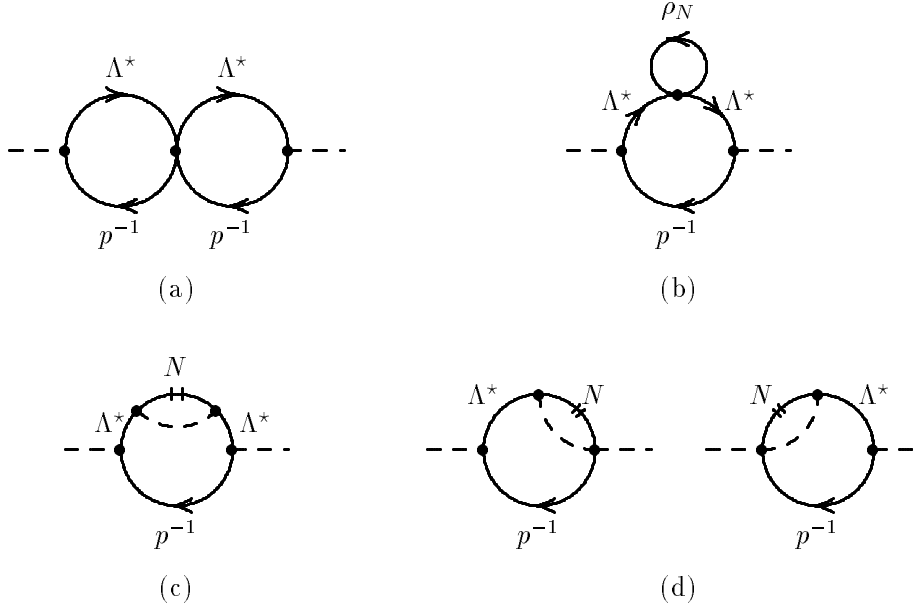,height=8cm}}
\caption[selfenergy] {The in-medium two-loop kaon self-energy
involving $\Lambda (1405)$. Figures a and b contain the constants
of the four-Fermi interaction and figures c and d are Pauli corrections}
\label{selfenergy}
\end{figure}

In order to confront kaonic atom data, we need to calculate the kaon
self-energy $\Pi$ in nuclei. The off-shell amplitude determined above
gives the so-called ``impulse" term
 \be
    \Pi^{imp}_K(\omega) &=& -\left( \rho_p  {\cal T}^{K^-p}_{free}(\omega)
        +\rho_n  {\cal T}^{K^-n}_{free}(\omega) \right)\label{self1}
    \ee
where ${\cal T}^{KN}$ is the off-shell s-wave KN transition matrix.
(The amplitude ${\cal T}^{KN}$ taken on-shell, {\it i.e.},
$\omega=M_K$, and the scattering length
$a^{KN}$ are related by $ a^{KN} = \frac{1}{4\pi (1+M_K/m_B)}
{\cal T}^{KN} $.)  Medium corrections to this ``impulse" term,
obtained from one-loop graphs by replacing the free-space
nucleon propagator by the in-medium propagator, shall be denoted as
\be
 - \left(\rho_p  \delta {\cal T}^{K^-p}_{\rho_N}(\omega)
        +\rho_n \delta {\cal T}^{K^-n}_{\rho_N}(\omega) \right).\label{deself}
\ee
These two terms (\ref{self1}) and (\ref{deself})
are completely determined by the parameters fixed above.
The new parameters of the four-Fermi interaction come into play in the
first two self-energy graphs of Fig.\ref{selfenergy}
(the last two graphs do not
involve four-Fermi interactions but enter at the same order; they are
free of unknown parameters),
 \be
    \Pi_{\Lambda^\star}(\omega) &=& - \frac{g_{\Lambda^\star}^2}{f^2}
        \left(\frac{\omega}{\omega+m_B-m_{\Lambda^\star}} \right)^2
    \left\{ C_{\Lambda^\star}^S \rho_p \left( \rho_n +\frac 12 \rho_p \right)
    -\frac 32 C_{\Lambda^\star}^T \rho_p^2 \right\}
    \nonumber\\
    \nonumber\\
    && + \frac{g_{\Lambda^\star}^2}{f^4} \rho_p
    \left(\frac{\omega}{\omega+m_B-m_{\Lambda^\star}} \right)
    \omega^2 \left\{  \left( 2 \Sigma_K^p (\omega) +\Sigma_K^n (\omega) \right)
    \vphantom{\frac 12} \right.
    \nonumber\\
    && \;\;\;\;\;\;\;\;\;\;\;\;\;\;\; \;\;\;\;\;\;\;\;\;\;\;\;\;\;\; \left.
     -g_{\Lambda^\star}^2
    \left(\frac{\omega}{\omega+m_B-m_{\Lambda^\star}} \right)
     \left( \Sigma_K^p (\omega) +\Sigma_K^n (\omega) \right)
    \right\}
    \label{pilambda}\ee
where ${g}_{\Lambda^\star}$ is the renormalized $KN\Lambda^\star$ coupling
constant determined in Lee {\etal}\cite{LJMR}
and $\Sigma_K^N (\omega)$ is a known integral that depends on proton and
neutron densities and $M_K$.
Note that while the second term of (\ref{pilambda})
gives repulsion corresponding to a Pauli quenching,
the first term can give either attraction or repulsion depending on
the sign of $(C_{\Lambda^\star}^S [\rho_n+\frac 12 \rho_p]-
\frac 32 C_{\Lambda^\star}^T\rho_p)$. For symmetric nuclear matter, only the
combination  $(C_{\Lambda^\star}^S - C_{\Lambda^\star}^T)$ enters in the
self-energy. This is an important element for kaonic atom.

The complete self-energy to in-medium two-loop order
is then
    \be
    \Pi_K(\omega) &=& -\left( \rho_p  {\cal T}^{K^-p}_{free}(\omega)
        +\rho_n  {\cal T}^{K^-n}_{free}(\omega) \right)
        - \left(\rho_p  \delta {\cal T}^{K^-p}_{\rho_N}(\omega)
        +\rho_n \delta {\cal T}^{K^-n}_{\rho_N}(\omega) \right)\nonumber\\
       && +\Pi_{\Lambda^\star}(\omega)\label{self2}.
    \ee

We now turn to fixing the constants of the four-Fermi interactions
based on the recent analysis of kaonic atoms by Friedman, Gal
and Batty \cite{kaonicatom}. For later purpose we shall parametrize
the proton and neutron densities by the proton fraction $x$
and the nucleon density $u=\rho/\rho_0$ as
    \be
    \rho_p = x\rho\; ,\;\; \rho_n =(1-x) \rho\; ,\;\; \rho = u\rho_0.
    \ee
Now Friedman {\etal}\cite{kaonicatom} found from their analysis that the
optical potential for the $K^-$ in medium has an attraction of the order of
    \be
    \Delta V\equiv M_K^\star-M_K
    \approx -(200\pm 20)\ \ {\rm  MeV}\ \ at \;\; u=0.97
    \ee
    with
    \be
M_K^\star \equiv\sqrt{M_K^2+\Pi_K}.
\ee
This implies approximately for $x=1/2$
    \be
     ( C_{\Lambda^\star}^S -C_{\Lambda^\star}^T) f^2\approx 20.
\label{cvalue}
    \ee
Friedman {\etal} \cite{kaonicatom} note that their ``nominal" optical
potential gives an attraction of order of 800 MeV when extrapolated
to three times the normal density. We show in Table 1 what our theory predicts
at higher densities than normal. At $u=3$, the net attraction is only about
1.7 times the one at $u=1$.

\vskip 0.5cm
    \begin{center}
    {\bf Table 1}\\
\vskip 0.3cm
    \parbox[t]{5.0in}{
     $K^-$ effective mass($M_K^*$) and the attraction
    ($\Delta V\equiv M_K^\star -M_K$ )
    in symmetric nuclear matter ($x=0.5$) as function
    of density $u$ in unit of MeV for
     $(C_{\Lambda^\star}^S-C_{\Lambda^\star}^T) f^2 =20$.
    }
    \end{center}
    $$
    \begin{array}{|c||c|c|}
    \hline
     u & M_K^* & \Delta V  \\
    \hline
    \hline
 0.5& 348.6& -146.4 \\ \hline
 1.0& 294.5& -200.5 \\ \hline
 1.5& 249.3& -245.8 \\ \hline
 2.0& 211.7& -283.3 \\ \hline
 2.5& 179.9& -315.1 \\ \hline
 3.0& 153.0& -342.0 \\ \hline
 3.5& 129.0& -366.0 \\ \hline
 4.0& 110.4& -384.6 \\ \hline
    \end{array}
    $$
\vskip 0.5cm

Equation (\ref{pilambda}) shows that for symmetric nuclear matter ($x=1/2$),
the combination $(C_{\Lambda^\star}^S +C_{\Lambda^\star}^T)$ does not enter
into the
self-energy formula. In order to extract it as needed for non-symmetric
system as in compact star matter, we need information for nuclei with
$x\neq 1/2$. This can be done from the results of Friedman {\etal}
by noting that our self-energy is nonlinear in $x$, so
\be
\frac{\del\Delta V}{\del x} (C_{\Lambda^\star}^S,\rho\approx \rho_0)|_{x=1/2}
\approx 400 \;  b_1/b_0 \ \ {\mbox MeV}
\ee
where $b_{0,1}$ are the constants given by Friedman {\etal}.  This relation
determines the coefficient $C_{\Lambda^\star}^S$. The result is shown in
Table 2 (first three columns).

Friedman {\etal}\cite{kaonicatom} find the acceptable value to be
$b_1/b_0= -0.56\pm 0.82$. But there is one point which needs to be discussed
in interpreting this number in the context of our theory.
The constant $C_{\Lambda^\star}^S$ shifts linearly the effective
in-medium mass of $\Lambda (1405)$, with the mass shift being given by
\be
\delta m_{\Lambda^\star}=\sum_{i=a,b}\delta
 \Sigma^{(i)}_{\Lambda^\star} (\omega=m_{\Lambda^\star}-m_B)
\ee
where
\be
\delta\Sigma_{\Lambda^\star}^{(a)} (\omega)
&=&-\frac{g_{\Lambda^\star}^2}{f^2}\omega^2\left(
\Sigma^p_K (\omega) +\Sigma^n_K (\omega)\right)\nonumber\\
\delta\Sigma_{\Lambda^\star}^{(b)} (\omega) &=&
-C_{\Lambda^\star}^S (\rho_p+\rho_n).
\ee
For nuclear matter density $u=1$ and $x=1/2$, the shift is
\be
\delta m_{\Lambda^\star} (u,x,y)\approx [62-150.3\times y] \ \ {\rm MeV}
\label{lambdamass}
\ee
with $y=C^S_{\Lambda^\star} f^2$. It seems highly unlikely that
the $\Lambda (1405)$ will be shifted by hundreds of MeV in nuclear matter.
This means that
$y$ must be of $O(1)$, and {\it not} $O(10)$.  For $y=0.41$
which corresponds to $b_1/b_0\approx -0.4$, there is no shift at normal matter
density.  We believe this is a reasonable value. In fact,
$y=0$ is also acceptable. It would be interesting to
measure the shift of $\Lambda (1405)$ to fix the constant
$C_{\Lambda^\star}^S$ more precisely although its precise magnitude
seems to matter only a little for kaonic atoms and as it turns out, negligibly
for kaon condensation.
\vskip 0.5cm

\centerline{\bf Table 2}
\begin{quotation}
\noindent Determination of $C_{\Lambda^\star}^S$ from the kaonic atom
data\cite{kaonicatom} and the critical density (obtained with the
constant so determined) for
kaon condensation for various forms of symmetry energy $F(u)$. $y=0.41$
corresponds to no $\Lambda (1405)$ mass shift in medium at the normal matter
density.
\end{quotation}
$$
\begin{array}{|c|r|r|c|c|c|}
\hline
 & & & \multicolumn{3}{c|}{ u_c } \\ \cline{4-6}
y=C_{\Lambda^\star}^S f^2 & \partial\Delta V/\partial x & b_1/b_0
& F(u)=\frac{2u^2}{1+u} & F(u)= u &  F(u)=\sqrt u \\
\hline
50  &  125.78 \MeV &  0.314 &      2.247 &      2.492 &      2.942 \\
40  &  65.11 \MeV &  0.163 &      2.320 &      2.572 &      3.051 \\
30  &  4.44 \MeV &  0.011 &      2.407 &      2.680 &      3.189 \\
20  &  -56.23 \MeV & -0.141 &      2.528 &      2.821 &      3.372 \\
10  & -116.91 \MeV & -0.292 &      2.696 &      3.033 &      3.645 \\
\hline \hline
0.41 & -175.089 \MeV & -0.438 &  2.958  & 3.391 &  4.159 \\
\hline \hline
0   & -177.58 \MeV & -0.444 &      2.973 &      3.414 &      4.195 \\
-10 & -238.25 \MeV & -0.596 &      3.564 &      4.741 & \sim 5.897 \\
\hline
\end{array}
$$
\vskip 0.5cm

Let us comment briefly on the role of multi-Fermion Lagrangians.
This will eliminate another red herring in the literature.
The Weinberg counting rule shows that the four-Fermi interactions
are suppressed by $O(Q^2)$ relative to the terms involving bilinears of
Fermi fields. In general $n$-Fermi interactions will be suppressed
by the same order relative to $(n-1)$-Fermi interactions. In considering
kaon condensation, what this means in conjunction with the
renormalization-group flow argument, is that $n$-Fermi interactions
with $n\geq 4$ are irrelevant in the RGE sense, and hence unimportant
for condensation. The situation with the kaonic atom data is a bit different.
While the strength of the four-Fermi interaction, $y$, is not important
(this can be seen in Lee {\etal}\cite{LBMR}, Table 3), its presence is
essential for the attraction that seems to be required. This is in contrast
to the kaon condensation which is driven by the ``mass flow" with four-Fermi
interactions being irrelevant in the RGE sense.

\section{Kaon Condensation}
We have now all the ingredients needed to calculate the critical density
for negatively charged kaon condensation in dense nuclear star matter.
For this, we will follow the procedure given in work of Thorsson, Prakash
and Lattimer (TPL)\cite{TPL}.
As argued by Brown, Kubodera and Rho\cite{BKR}, we need not consider pions
when electrons with high chemical potential can trigger condensation through
the process $e^-\rightarrow K^- \nu_e$. Thus we can focus on the spatially
uniform condensate
    \be
    \langle K^-\rangle =v_K e^{-i\mu t}.
    \ee
The energy density $\tilde\epsilon$ -- which is related to the
effective potential in the standard way -- is given by,
    \be
    \tilde \epsilon (u,x,\mu, v_K) &=& \frac 35 E_F^{(0)} u^{\frac 53} \rho_0
        +V(u) +u\rho_0 (1-2x)^2 S(u) \nonumber\\
    &&-[\mu^2 -M_K^2 -\Pi_K (\mu,u,x)]
        v_K^2+ \sum_{n\ge 2} a_n(\mu,u,x) v_K^n \nonumber\\
    && +\mu u\rho_0 x +\tilde\epsilon_e +\theta(|\mu|-m_\mu)\tilde \epsilon_\mu
    \label{effen}\ee
where $E_F^{(0)}=\left( p_F^{(0)}\right)^2/2m_B$ and
$p_F^{(0)}=(3\pi^2\rho_0 /2)^{\frac 13}$ are, respectively,
 Fermi energy and momentum at nuclear density. The
$V(u)$ is a potential for symmetric nuclear matter
as described by Prakash {\etal}\cite{PAL}
which is presumably subsumed in contact four-Fermi
interactions (and one-pion-exchange -- nonlocal -- interaction)
in the non-strange sector as mentioned above. It will affect
the equation of
state in the condensed phase but not the critical density, so we will
drop it from now on. The nuclear symmetry energy $S(u)$ -- also
subsumed in four-Fermi interactions in the non-strange sector -- does
play a role as we know from Prakash {\etal}\cite{PAL}:
Protons enter to neutralize the
charge of condensing $K^-$'s making the resulting compact star
``nuclear" rather than neutron star as one learns in standard astrophysics
textbooks. We take the form advocated by Prakash {\etal}\cite{PAL}
    \be
    S(u) &=& \left(2^{\frac 23}-1\right) \frac 35 E_F^{(0)}
        \left(u^{\frac 23} -F(u) \right) +S_0 F(u)
    \ee
where $F(u)$ is the potential contributions to the symmetry energy and
$S_0 \simeq 30 MeV$ is the bulk symmetry energy parameter.
We use three different forms of $F(u)$\cite{PAL}
    \be
    F(u)=u\;,\;\; F(u) =\frac{2u^2}{1+u}\;,\;\; F(u)=\sqrt u.
    \label{SE}
    \ee
The contributions of the filled Fermi seas of electrons and muons
are\cite{TPL}
    \be
    \tilde \epsilon_e &=& -\frac{\mu^4}{12\pi^2} \nonumber\\
    \tilde \epsilon_\mu &=& \epsilon_\mu -\mu \rho_\mu
    = \frac{m_\mu^4}{8\pi^2}\left((2t^2+1) t\sqrt{t^2+1}
    -\ln(t^2+\sqrt{t^2+1}
    \right) -\mu \frac{p_{F_\mu}^3}{3\pi^2}
    \ee
where $p_{F_\mu} =\sqrt{\mu^2-m_\mu^2}$ is the Fermi momentum and $t=p_{F_\mu}
/m_\mu$.

The ground-state energy prior to kaon condensation
is then obtained by extremizing the energy density $\tilde\epsilon$
with respect to  $x$, $\mu$ and $v_K$:
    \be
    \left. \frac{\partial\epsilon}{\partial x}\right |_{v_K=0}=0 \;,\;\;
    \left. \frac{\partial\epsilon}{\partial \mu}\right |_{v_K=0}=0 \;,\;\;
    \left. \frac{\partial\epsilon}{\partial v_K^2}\right |_{v_K=0}=0
    \ee
from which we obtain three equations corresponding, respectively,
to beta equilibrium, charge neutrality and dispersion relation.
The critical density so obtained is given for  three different $F(u)$'s
in Table 2. The result is
\be
2< u_c\lsim 4.\label{ineq}
\ee

\section{Discussion}
We note that the largest sensitivity is associated with the part that is
not controlled by chiral symmetry, namely the density dependence of the
symmetry energy function $F(u)$. This uncertainty reflects the part of
interaction that is not directly given  by chiral Lagrangians, that is,
the part leading to normal nuclear matter. This is the major short-coming of
our calculation.

Related to this issue is BR scaling. As we argued, were we able to
derive nuclear matter from effective chiral Lagrangians, we would have
parameters of the theory determined at that point reflecting
the background around which fluctuations are to be made.
The BR scaling was proposed in that spirit but with a rather strong
assumption: That a sigma model governs dynamics in medium as in free space
with only coupling constants and masses scaled a function of density.
Up to date, no derivation of this scaling from basic principles
has been made. In this sense, we might consider it as a conjecture.
Suppose we apply
BR scaling. The only way the procedure can make sense is to apply the
scaling argument to the tree order terms, but not to the loop corrections.
The result of this procedure is significant in that the critical density
is brought down in an intuitively plausible way to about $u_c\sim 2$,
with very little dependence on parameters, loop
corrections and multi-Fermi interactions. Thus slightly modified from
(\ref{ineq}), we arrive at the announced result
\be
2\lsim  u_c\lsim 4.\label{ineqq}
\ee

Whether or not the Bethe-Brown scenario\cite{BB} of compact star formation
is supported by the chiral Lagrangian approach will have to await the
calculation of the equation of state at in-medium two-loop order,
which is in progress. Our conjecture is that to the extent that our
work confirms the original Kaplan-Nelson calculation\cite{KN},
the compact star
properties calculated previously at the tree level\cite{TPL} would
come out qualitatively unmodified in the higher-order chiral perturbation
theory.

\section*{Acknowledgments}
We would like to thank G.E. Brown, H.K. Lee, D.-P. Min and
S.-J. Sin for discussions. The work of CHL is supported in part by the
Korea Science and Engineering Foundation through the CTP of SNU and in
part by the Korea Ministry of Education under Grant No. BSRI-94-2418.

\end{document}